\documentclass[11pt]{amsart}

\usepackage{amsmath,amsfonts}
\usepackage{latexsym,amssymb}
\usepackage{amsthm}
\usepackage{bm}
\usepackage{graphicx}

\newcommand{\dd}{\mathtt{d}}

\numberwithin{equation}{section}

\begin{document}

\title[] {Can exist a function, that transforms electromagnetic potentials from one to other gauge?}
\author{Vladimir~Onoochin}

\begin{abstract}
In this article, it is analyzed a problem of existence of a function which is able to transform electromagnetic potentials defined in one gauge to corresponding potentials defined in the other gauge. It is shown that such a function cannot exist.
\end{abstract}
\maketitle

\section{Introduction}

Despite electromagnetic potentials are treated within the classical electrodynamics as a mathematical tool to calculate the physical quantities, the EM fields, computation of these potentials is mostly forced procedure. The Maxwell equations are solved by means of the wave equations for the scalar $\varphi $ and vector ${\bf A}$ potentials. As it had been shown by Lorenz still in 1867, solutions of these wave equations plus the condition connecting the potentials (in the Gauss units),  
\begin{equation}
\bm{\nabla}\cdot{\bf A}+\frac{1}{c}\frac{\partial \varphi }{\partial t}=0\,, \label{lg}
\end{equation}
fully correspond to the Maxwell equations~\cite{OR}.

According to~\cite{JO}, Lorentz re-opened the gauge condition in 1904 by writing that in order to have the potentials satisfy the ordinary wave equations they must be related by~(\ref{lg}). He discusses the arbitrariness in the potentials and then states 'every other admissible pair ${\bf A}'$ and $\varphi '$ can be related to the first pair ${\bf A}'$ and $\varphi '$ via the transformations
\begin{equation}
\begin{split}
 \varphi' = \varphi -\frac{1}{c}{\frac {\partial {\chi_L }}{\partial t}}\,, \\
 {\bf A}' = {\bf A} +\bm{\nabla} \chi_L\,, 
 \end{split}\label{Av}
\end{equation} 
where the function $\chi_L({\bf r},t)$ can be determined by subjecting ${\bf A}'$ and $\varphi '$ to the condition~\eqref{lg}~(Note 5 of~\cite{L})
\begin{equation}
\nabla\cdot{\bf A}'+\frac{1}{c}\frac{\partial \varphi ' }{\partial t}=0\,,\quad\to\quad\nabla^2\chi_L-\frac{1}{c^2}\frac{\partial^2\chi_L}{\partial t^2} =
\nabla\cdot{\bf A}+\frac{1}{c}\frac{\partial \varphi }{\partial t}\,.
\end{equation}
So the gauge function in the Lorenz gauge obeys the homogeneous wave equation.

After these results of Lorentz, the function $\chi_L$ is called the {\it gauge function}. Let us note that Lorentz introduced transformations (\ref{Av}) for the potentials those obey the wave equations (p. 157 of~\cite{Lor1904}), or defined in the Lorenz gauge.

Essential property of transformations~(\ref{Av}) is that both sets of the potentials, $\varphi,\,{\bf A}$ and  $\varphi',\,{\bf A}'$, give the same values of the EM fields since the derivatives of the gauge function mutually compensate each other in the Maxwell equations.

Later these transformations have been extended to other gauges, first of all, the Coulomb gauge. The latter is associated with Maxwell who derived the wave equation for the magnetic field using the condition (in modern notation)
\begin{equation}
\bm{\nabla}\cdot \mathbf{A}=0\,. \label{cg}
\end{equation}
that was later called as the {\it Coulomb gauge}. Let us apply the transformations of the same form as for the gauge function in the Lorenz gauge,
\begin{equation}
 {\bf A}_C' = {\bf A}_C +\nabla \chi_C\,, \label{A}
\end{equation}
where ${\bf A}_C,\,{\bf A}_C'$ are the vector potentials in the Coulomb gauge. Using the gauge condition~(\ref{cg}), one has
\begin{equation}
 \bm{\nabla}\cdot{\bf A}_C' = \bm{\nabla}\cdot{\bf A}_C +\nabla^2 \chi_C=0\,, 
\end{equation}
Since both $\bm{\nabla}\cdot{\bf A}_C'$ and $\bm{\nabla}\cdot{\bf A}_C$ should be equal to zero in this gauge, one obtains that the Coulomb gauge function obeys the Poisson equation $\nabla^2\chi_C=0$.

Although most problems in electrodynamics are solved in the Lorenz gauge, where the wave equations for the potentials are separated, there are some problems that are solved in another gauge. For example, due to the difficulties of quantizing scalar photons, some problems of quantum electrodynamics are solved in the Coulomb gauge, where the scalar potential is considered as a static non-quantizable quantity. The use of different gauges raises the question, are the electromagnetic fields calculated from the potentials determined in different gauges equal? The study of this problem raises another question: is it possible to find a function that transforms electromagnetic potentials defined in one gauge into electromagnetic potentials defined in another gauge. It will be shown below that such a function cannot exist in the general case.

Since the most used gauges are the Lorenz and Coulomb ones, it is appropriate to find a function that transforms the potentials of these gauges.

\section{The condition of the gauge function existence.}

If one intends to seek such a gauge function, first of all one should be sure that the derivatives of this function do not enter into the Maxwell equations. So it is reasonable to choose the same form for transformations of the potentials as it was introduced by Lorentz, or
\begin{eqnarray}
 \varphi_C - \varphi_L =-\frac{1}{c}{\dfrac {\partial {\chi }}{\partial t}}\,,\label{gau1} \\
 {\bf A}_C- {\bf A}_L =\bm{\nabla} \chi\,, \label{gau2}
\end{eqnarray} 
where the indices "C" and "L" refer to the potentials defined in the Coulomb and Lorentz gauges, and $\chi$ is the required function. This function must satisfy the main requirement for the gauge function, namely, in the Maxwell equations, the derivatives of $\chi$ will eliminate each other and the gauge function cannot be included in the Maxwell equations.

At the same time, fulfillment of Eqs.~\eqref{gau1} and~\eqref{gau2} provides the equivalence of the EM fields calculated in both gauges. Actually,
\begin{equation*}\label{gaug}
\begin{split}
 \bm{\nabla}\varphi_C - \bm{\nabla}\varphi_L =-\frac{1}{c}\bm{\nabla}{\frac {\partial {\chi }}{\partial t}}\,, \\
 \frac{1}{c}\frac{\partial {\bf A}_C}{\partial t}-\frac{1}{c}\frac{\partial {\bf A}_L}{\partial t} =
\frac{1}{c}\bm{\nabla}{\frac {\partial {\chi }}{\partial t}}\,.
 \end{split}
\end{equation*}
Term-by-term summation of the above equations gives
\begin{equation*}
 \bm{\nabla}\varphi_C +  \frac{1}{c}\frac{\partial {\bf A}_C}{\partial t} -
  \bm{\nabla}\varphi_L -  \frac{1}{c}\frac{\partial {\bf A}_L}{\partial t} =0\,\,\to\,\,{\bf E}_L-{\bf E}_C=0\,.\label{gaugs}
\end{equation*}
In order for the above equation to be valid, one needs to find the functions $\chi$, or at least show that this function exists. To realize this statement, let us analyze the restrictions that are imposed on $\chi$ by Eqs.~\eqref{gau1} and ~\eqref{gau2}.

Calculation of the divergence of both parts of Eq.~\eqref{gau1} gives,
\begin{equation}
\bm{\nabla}\cdot{\bf A}_C-\bm{\nabla}\cdot{\bf A}_L =\bm{\nabla}\cdot\bm{\nabla} \chi\,\,\quad{\rm or}\,\, \quad
-\bm{\nabla}\cdot{\bf A}_L =\nabla^2 \chi\,,
\end{equation}
and after application of the Lorenz gauge condition to the {\it lhs} of the above equation, one has
\begin{equation}
\frac{1}{c}\Dot{\varphi_L}= \nabla^2 \chi\,. \label{2}
\end{equation}
The gauge function can be reconstructed from Eq.~(\ref{2}), so one obtains,
\begin{equation}
\chi=\frac{1}{4\pi}\int \frac{ \nabla^2 \chi({\bf r}')}{|{\bf r}-{\bf r}' |}\dd ^3r'= \frac{1}{4\pi c}\int \frac{\Dot{\varphi_L}({\bf r}',t)}{|{\bf r}-{\bf r}' |}\dd ^3r' \,.\label{Chi}
\end{equation}
The last integral in~\eqref{Chi} diverges as $\ln r$ since $\Dot{\varphi_L}\simeq 1/r^2$ but let us skip this defect for a while.

Since the difference ${\bf A}_C-{\bf A}_L$ is given by the solution of the wave equation with the {\it rhs} equal to  $\dfrac{1}{c}\bm{\nabla}\dfrac{\partial \varphi _C}{\partial t}$ (Eq.~(3.3) of~\cite{Onoo}), and ${\bf A}_C-{\bf A}_L=\bm{\nabla}\chi$, one has,
\begin{equation}
\nabla\chi({\bf r},t)=\frac{1}{4 \pi c}\int\frac{[\bm{\nabla}\Dot{\varphi}_C]_{ret}}{|{\bf r}-{\bf r}'|}\dd ^3r' \,.
\label{Ch}
\end{equation}
Thus, gradient of the gauge function should be equal to the integral in {\it rhs} of~(\ref{Chi}) and at the same time this gradient should be equal to the integral in {\it rhs} of~(\ref{Ch}). Therefore, these integrals should be equal one other,
\begin{equation*}
\frac{1}{4 \pi c}\int\frac{[\bm{\nabla}_{r'}\Dot{\varphi}_C({\bf r}',t']_{ret}}{|{\bf r}-{\bf r}'|}\dd ^3r'  = 
\frac{1}{4 \pi c}\bm{\nabla}_r\int \frac{\Dot{\varphi_L}({\bf r}')}{|{\bf r}-{\bf r}' |}\dd ^3r'\,. \label{finn}
\end{equation*}
where the argument $t'$ in $\varphi_C$ corresponds to the retarded time -- actually, the change $t'\,\to\,t_{r}=t-|{\bf r}-{\bf r}'|/c$ should be made after preforming all operations on $\Phi_C$

Typically,  the scalar potential in the Lorenz gauge is expressed via the retarded time (for example, the Li\'enard-Wiechert potentials). However, if $\varphi_L$ is expressed in the present time coordinates, Eq.~(\ref{fin}) is simplified. Since in this case $\Dot{\varphi_L}$ is function only of $r'$ and $t$,
\[
\bm{\nabla}_r\int \frac{\Dot{\varphi_L}({\bf r}',t)}{|{\bf r}-{\bf r}' |}\dd ^3r' =\int \frac{\bm{\nabla}_{r'}\Dot{\varphi_L}({\bf r}',t)}{|{\bf r}-{\bf r}' |}\dd ^3r' \,,
\]
and
\begin{equation}
\frac{1}{4 \pi c}\int\frac{[\bm{\nabla}_{r'}\Dot{\varphi}_C({\bf r}',t]_{ret}}{|{\bf r}-{\bf r}'|}\dd ^3r'  = 
\frac{1}{4 \pi c}\int \frac{\bm{\nabla}_{r'}\Dot{\varphi_L}({\bf r}',t)}{|{\bf r}-{\bf r}' |}\dd ^3r'\,. \label{fin}
\end{equation}
The above relation must be fulfilled with necessity. If it is not fulfilled, it means that there is no function providing the transformation of the potentials from the Coulomb to Lorenz gauge, and otherwise.

It is difficult to verify this relation, since the integrals in it are too cumbersome. However, these integrals can be computed in simplest case of uniform motion of a single classical charge. Let us show how the integrals of Eq.~\eqref{fin} can be computed.

\section{Computation of the integrals.}

Verification of the fulfillment of relation~\eqref{fin}, for example, for the $x$ component of $\bm{\nabla}\Phi$, can be made by consideration of potentials created by a classical charge which is at rest until $t=0$, and then begins to move with constant velocity $v$ along the $x$ axis. 

In this case, only the $x$ component of integrals can have non-zero values. Due to symmetry, the $y$ and $z$ components should be equal to zero.

If the charge starts its motion from the point $r=0$, the scalar potentials in two gauges will be
\begin{equation*}
\Phi_C({\bf r},t)=\left\{
\begin{array}{ccc}
\dfrac{q\Theta(-t)}{\sqrt{x^2+y^2+z^2}}\,;\\
\dfrac{q\Theta(t)}{\sqrt{(x-vt)^2+y^2+z^2}}\,;
\end{array}\right.;\,\,
\Phi_L({\bf r},t)=\left\{
\begin{array}{ccc}
\dfrac{q\Theta(-t)}{\sqrt{x^2+y^2+z^2}}\,;\\
\dfrac{q\Theta(t)}{\sqrt{(x-vt)^2+[1-(v/c)^2][y^2+z^2]}}\,;
\end{array}\right.
\end{equation*}
where $q$ is a charge of the particle and $\Theta(t)$ is the step function. Calculation of $\partial_x\Dot{\Phi}$ for these potentials gives
\begin{equation}
\begin{split}
\partial_x\Dot{\Phi}_C({\bf r},t)=\dfrac{vq\Theta(t)\left[y^2+z^2-2(x-vt)^2 \right]}{\left[(x-vt)^2+y^2+z^2\right]^{5/2}}\,,\\
\partial_x\Dot{\Phi}_L({\bf r},t)=\dfrac{vq\Theta(t)\left[[1-(v/c)^2](y^2+z^2)-2(x-vt)^2 \right]}{\left[(x-vt)^2+[1-(v/c)^2](y^2+z^2)\right]^{5/2}}\,.
\end{split}
\end{equation}
Here, it should be noted some feature of establishing the scalar potentials in space after change of motion of the charge at $t=0$. If $\Phi_C({\bf r},t)$ is establishing in all space instantaneously, it is not the case for $\Phi_L({\bf r},t)$.

'Switching on' the potential of moving charge at a point $O$ (fig 1) with ${\bf r}=0$ at the instant $t=0$ means that a spherical 'wave of establishing' of this potential in space begins to extend from $O$ and at an instant $t$ a front of this wave will be at $R=ct$.

Let us choose the origin of the coordinate system is located at $r'=0$ and a value of the external variable $r=0$.  Then in spherical coordinates, $r',\,\theta,\,\phi$ the region of integration is the same for both integrals (fig. 1). For $\Phi_L$, this region is determined by condition that at the interval $[0;t]$ the potential of the moving charge can be established inside the sphere with the radius $r'=R=ct$. For $\Phi_C$, this region is determined by the argument of the $\Theta$ function, $t-r'/c>0$.
\begin{center}
\begin{figure}[h]%
\includegraphics[bb = 0 0 629 381, scale=0.6]{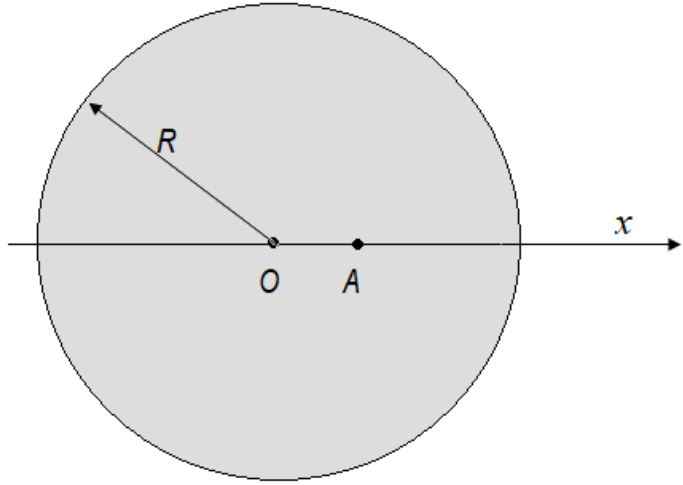}
\caption{The sphere of integration. The radius of the sphere $R=ct$. The charge begins its motion from a point $O(0,0,0)$, and its position at the instant $t$ is at the point $A(vt,0,0)$, so $OA=vt$.}\label{fig1}
\end{figure}
\end{center}
The symbol $[\,\,\,]_{ret}$ in Eq.~\eqref{fin} means that the nominator in its {\it lhs}  depends on the retarded time, or $t_r=t-r'/c$. Then in the spherical coordinates, the integrals are written as~\footnote{Due to axial symmetry (the $x$ axis is the axis of symmetry), integration over $\phi$ gives $2\pi$.}
\begin{eqnarray}\nonumber
\frac{q}{2 c} \int \limits_0^{ct} r'\dd r'\int\limits_0^{\pi}\dfrac{v\left[r'^2(1-3\cos^2\theta)+4vr't_r\cos\theta-2v^2t_r^2  \right]\sin\theta\dd\theta }
{\left[r'^2+v^2t_r^2+2vr't_r\cos\theta  \right]^{5/2} }=\\
=\frac{q}{2 c}\int \limits_0^{ct} r'\dd r'\int\limits_0^{\pi}\dfrac{v\left[r'^2(1-v^2/c^2)(1-\cos^2\theta)
-2(r'\cos\theta - vt)^2  \right] \sin\theta\dd\theta }
{\left[r'^2(1-v^2/c^2)(1-\cos^2\theta) +(r'\cos\theta - vt)^2  \right]^{5/2} }\,, \label{Ints}
\end{eqnarray}
Both integrands contain the regular and singular terms, or the terms which make the above integrals belong to  improper ones. The latter appear due to the second derivative of the scalar potential, and special procedure has been developed for processing divergent (singular) terms (Ch.~IV.5.5 of~\cite{Tikh}). Thus, computation of the integrals with these terms gives
\[
I_{C,{\rm sing}}=\frac{q}{3c}\int \frac{1}{\sqrt{x'^2+y'^2+z'^2}}\dfrac{\delta(x'-\tilde {x})\delta(y')\delta(z')}{\dfrac{c+v}{c}}\dfrac{\dd x'\dd y'\dd z'}{\tilde{x}-
\dfrac{cvt}{c+v}}=\frac{q}{3ct}\,,
\]
\[
I_{L,{\rm sing}}=\frac{q }{3c}\int \frac{\delta(x'-vt)\delta(y')\delta(z')}{\sqrt{x'^2+y'^2+z'^2}}\dd x'\dd y'\dd z'=\frac{q}{3ct}\,.
\]
So account of the singular terms of the integrands is equal for both integrals. Now it is necessary to compute accounts of  regular terms. After change of variable $\cos\theta = \xi$, the first of integrals~\eqref{Ints} becomes,
\[
\frac{q}{2 c}\int \limits_0^{ct} r'\dd r'\int\limits_{-1}^{1}\dfrac{v\left[r'^2(1-3\xi)+4vr't_r\xi-2v^2t_r^2  \right]\dd\xi }
{\left[r'^2+v^2t_r^2+2vr't_r\xi  \right]^{5/2} }=-\frac{q}{2 c}\int \limits_0^{ct} r'\dd r'\frac{2(1-{\rm sgn}(r'-vt_r))}
{v^2t^3}
\]
where ${\rm sgn}(z)$ is a signum function of the real argument $z$. Thus, calculation of this integral with respect to $r'$ gives
\begin{equation}
-\frac{q}{2 c}\int \limits_0^{ct} r'\dd r'\frac{2\left(1-{\rm sgn}\left[r'-v(t-r'/c)\right]\right)}
{v^2t^3} =-\frac{2q}{v^2\,c}\int\limits_0^{vct/(c+v)}\frac{r'\dd r'}{(t-r'/c)^3}=-\frac{q}{ct}\,.\label{r1}
\end{equation}
Calculation of the second integral in~\eqref{Ints} with respect to $\xi$ by means of {\tt Mathematica} and then with respect to $r'$ gives
\begin{equation}\label{r2}
\begin{split}
&\frac{q}{2 c}\int \limits_0^{ct} r'\dd r'\int\limits_{-1}^{1}\dfrac{v\left[r'^2(1-v^2/c^2)(1-\xi^2)
-2(r'\xi - vt)^2  \right] \dd\xi }
{\left[r'^2(1-v^2/c^2)(1-\xi^2) +(r'\xi - vt)^2  \right]^{5/2} }=\\
&=-\frac{q}{2 c}\int \limits_0^{ct} \dd r'\frac{4c^4tr'\Theta(vt-r')}{v^2(r'^2-c^2t^2)^2}=-\frac{q}{\left(1-{v^2}/{c^2}\right)ct}\,.
\end{split}
\end{equation}
Since the results of integrations of Eqs.~\eqref{r1} and~\eqref{r2}  are not equal each other, and since the gauge function should correspond to both these expressions, such a gauge  function cannot exist in the general case.

\section{Conclusions.}

In this work it is considered a problem of existence of so-called gauge function that is able to transform the electromagnetic potentials defined in one gauge to the corresponding potentials defined in another gauge. Except the only requirement caused by the '{\it gauge invariance}', namely, that the Maxwell equations cannot contain any gauge function including its derivatives, no other specification of the form of the gauge function is used in the presented analysis.

Meanwhile, the requirement of the {\it gauge invariance}, which determines the relationship of potentials with derivatives of $\chi$--function, is sufficiently strong to determine that this function must be defined through scalar potentials. As a result, it turns out that this function is expressed in terms of the scalar potential in the Coulomb gauge, Eq.~\eqref{Ch} and at the same time it is expressed through the scalar potential in the Lorenz gauge, Eq.~\eqref{Chi}. The compatibility of these two conditions leads to Eq.~\eqref{fin}, which means that if this equation holds, then the gauge function can be found from any of Eqs.~\eqref{Ch} and~\eqref{Chi}. But if Eq.~\eqref{fin} does not hold, then the gauge function cannot exist.

However, there is one obstacle to verify a validity of Eq.~\eqref{fin} in the general case. The reason for this is that for such a verifying one needs to know the explicit form of the potentials of the charge in an arbitrary motion, written in the present time variables\footnote{$\Phi_L$ in Eq.~\eqref{fin} is a function of $t$ but not of $t_{ret}$.}. This explicit form is known, the expressions for the Li\'enard--Wiechert potentials. But these expressions are  written in terms of retarded variables. To check the fulfillment of Eq.~\eqref{fin}, it is necessary to know the expressions for the scalar potentials written in the present time variables.

But even if Eq.~\eqref{fin} does not hold in one particular case, this is sufficient to conclude that the needed connection between $\Phi_C$ and $\Phi_L$ does not exist. Such a special case is considered in Sec. III of this work and it is shown that connection between the scalar potentials defined in two gauges actually does not exist. Thus, even whether a problem of divergence of the integral~\eqref{Chi}, detrmining the gauge function, is ignored, the obtained result allows us to conclude that the required connection is generally invalid and, therefore, there is no gauge function that provides the transformation of electromagnetic potentials.

\end{document}